\documentclass[aps,pra,amssymb,twocolumn,superscriptaddress,showpacs]{revtex4}

\usepackage{graphicx}
\usepackage{dcolumn}
\usepackage{bm}
\usepackage{color}

\begin{document}

\newcommand{\beq}{\begin{equation}}
\newcommand{\eeq}{\end{equation}}
\newcommand{\barr}{\begin{eqnarray}}
\newcommand{\earr}{\end{eqnarray}}

\newcommand{\andy}[1]{ }

\newcommand{\bmsub}[1]{\mbox{\boldmath\scriptsize $#1$}}

\def\bra#1{\langle #1 |}
\def\ket#1{| #1 \rangle}
\def\sinc{\mathop{\text{sinc}}\nolimits}
\def\cV{\mathcal{V}}
\def\cH{\mathcal{H}}
\def\cT{\mathcal{T}}
\renewcommand{\Re}{\mathop{\text{Re}}\nolimits}
\newcommand{\tr}{\mathop{\text{Tr}}\nolimits}

\newcommand{\REV}[1]{\textbf{\color{red}#1}}
\newcommand{\BLUE}[1]{\textbf{\color{blue}#1}}
\newcommand{\GREEN}[1]{\textbf{\color{green}#1}}

\title{Probability density function characterization of multipartite entanglement}

\author{P. Facchi}
\affiliation{Dipartimento di Matematica, Universit\`a di Bari,
        I-70125  Bari, Italy}
\affiliation{INFN, Sezione di Bari, I-70126 Bari, Italy}
\author{G. Florio} \affiliation{Dipartimento di Fisica,
Universit\`a di Bari,
        I-70126  Bari, Italy}
\affiliation{INFN, Sezione di Bari, I-70126 Bari, Italy}
\author{S. Pascazio} \affiliation{Dipartimento di Fisica,
Universit\`a di Bari,
        I-70126  Bari, Italy}
\affiliation{INFN, Sezione di Bari, I-70126 Bari, Italy}

\date{\today}

\begin{abstract}
We propose a method to characterize and quantify multipartite
entanglement for pure states. The method hinges upon the study of
the probability density function of bipartite entanglement and is
tested on an ensemble of qubits in a variety of situations. This
characterization is also compared to several measures of
multipartite entanglement.
\end{abstract}


\maketitle

\section{Introduction}
\label{sec:Introduction}

Entanglement is one of the most intriguing features of quantum
mechanics. Although it is widely used in quantum communication and
information processing and plays a key role in quantum computation,
it is not fully understood. It is deeply rooted into the linearity
of quantum theory and in the superposition principle and basically
consists (for pure states) in the impossibility of factorizing the
state of the total system in terms of states of its constituents.

The quantification of entanglement is an open and challenging
problem. It is possible to give a good definition of
\emph{bipartite} entanglement in terms of the von Neumann entropy
and the entanglement of formation \cite{woot}. The problem of
defining \emph{multipartite} entanglement is more difficult
\cite{druss} and no unique definition exists:
different measures capture in general different aspects of the
problem \cite{multipart}. Attempts to quantify the degree of quantum
entanglement are usually formulated in terms of its behavior under
local operations/actions that can be performed on different
(possibly remote) parts of the total system. Some recent work has
focused on clarifying the dependence of entanglement on disorder and
its interplay with chaos \cite{entvschaos,SC}, or its behavior
across a phase transition \cite{QPT,tognetti}.

The work described here is motivated by the observation that as the
size of the system increases, the number of measures (i.e.\ real
numbers) needed to quantify multipartite entanglement grows
exponentially. A good definition of multipartite entanglement should
therefore hinge upon some statistical information about the system.
We shall look at the distribution of the purity of a subsystem over
all possible bipartitions of the total system. As a characterization
of multipartite entanglement we will not take a single real
\emph{number}, but rather a whole \emph{function}: the probability
density of bipartite entanglement between two parts of the total
system. The idea that complicated phenomena cannot be
``summarized" in a single (or a few) number(s) stems from studies
on complex systems \cite{parisi} and has been considered also in
the context of quantum entanglement \cite{MMSZ}. In a few words,
we expect that multipartite entanglement be large when bipartite
entanglement is large \emph{and} does not depend on the
bipartition, namely when its probability density is a narrow
function centered at a large value. This characterization of
entanglement will be tested on several classes of states and will
be compared with several measures of multipartite entanglement.

\section{The system}
\label{sec:syst}

We shall focus on a collection of $n$ qubits. The dimension of the
Hilbert space is $N=2^n$ and the two partitions $A$ and $B$ are made
up of $n_A$ and $n_B$ spins ($n_A+n_B=n$), respectively, where the
total Hilbert space reads
$\mathcal{H}=\mathcal{H}_A\otimes\mathcal{H}_B$ and the Hilbert
spaces $\mathcal{H}_A$ and $\mathcal{H}_B$ have dimensions
$N_A=2^{n_A}$ and $N_B=2^{n_B}$, respectively ($N_AN_B=N$). We shall
consider only pure states
\begin{equation}
|\psi\rangle = \sum_{k=0}^{N-1} z_k |k\rangle ,
\label{eq:genrandomx}
\end{equation}
where $|k\rangle=|j_A\rangle\otimes|l_B\rangle$, with a bijection
between $k$ and $(j_A,l_B)$, $0\le j_A \le N_A-1$ and $0\le l_B
\le N_B-1$. As a measure of bipartite entanglement between $A$ and $B$
we consider the participation number
\begin{equation}\label{eq:NAB}
N_{AB}=\pi_{AB}^{-1}, \quad \pi_{AB}=\tr_A \rho_A^2, \quad
\rho_A=\tr_B \rho,
\end{equation}
where $\rho=|\psi\rangle\langle\psi|$, and $\tr_A$ ($\tr_B$) is the
partial trace over the degrees of freedom of subsystem $A$ ($B$).
$N_{AB}$ can be viewed as the relevant number of terms in the
Schmidt decomposition of $|\psi\rangle$
\cite{eberly}. The quantity $n_{AB}= \log_2 N_{AB}$ represents the
effective number of entangled spins. Clearly, for a completely
separable state, $\tr_A \rho_A^2=1$ for all possible bipartitions,
yielding $N_{AB}=1$ and $n_{AB}=0$. In this sense the participation
number can distinguish between entangled and separable states.
Moreover $\pi_{AB}$ is directly related to the linear entropy
$S_L=1-\pi_{AB}$, that is an entanglement monotone, i.e.\ it is non
increasing under local operations
\cite{linentropy} and classical communication. In general, the
quantity $N_{AB}$ will depend on the bipartition, as in general
entanglement will be distributed in a different way among all
possible bipartitions. Therefore, its distribution $p(N_{AB})$
will yield information about multipartite entanglement: its mean
will be a measure of the amount of entanglement in the system,
while its variance will measure how well such entanglement is
distributed, a smaller variance corresponding to a higher
insensitivity to the particular choice of the partition.

We will show that for a large class of pure states, statistically
sampled over the unit sphere, $p(N_{AB})$ is very narrow and has a
very weak dependence on the bipartition: thus entanglement is
uniformly distributed among all possible bipartitions. Moreover,
$p(N_{AB})$ will be centered at a large value. These are both
signatures of a very high degree of \emph{multipartite}
entanglement.

By plugging (\ref{eq:genrandomx}) into (\ref{eq:NAB}) one gets
\begin{equation}\label{eq:piAB}
\pi_{AB}=\sum_{j,j'=0}^{N_A-1}\,\sum_{l,l'=0}^{N_B-1}z_{j l} \bar
z_{j' l} z_{j' l'} \bar z_{j l'}.
\end{equation}
We note that $\pi_{AB}=\tr_A \rho_A^2= \tr_B
\rho_B^2$ and $1/N_A \leq \tr_A \rho_A^2\leq 1$, with the
minimum (maximum) value attained for a completely mixed (pure) state
$\rho_A$. Therefore,
\begin{equation}\label{eq:propNAB}
1\leq N_{AB}=N_{BA}\leq \min\{N_A,N_B\}.
\end{equation}
A larger value of $N_{AB}$ corresponds to a more entangled
bipartition $(A,B)$, the maximum value being attainable for a
\emph{balanced} bipartition, i.e.\ when $n_A = [n/2]$ (and
$n_B=[(n+1)/2]$), where $[x]$ is the integer part of the real $x$,
that is the largest integer not exceeding $x$, and the maximum
possible entanglement is $N_{AB} = N_A= 2^{n_A}=\sqrt{N}$
($=\sqrt{N/2}$) for an even (odd) number of qubits. As anticipated,
as a characterization of multipartite entanglement we will consider
the distribution of $N_{AB}$ over all possible balanced
bipartitions.

\section{Measuring multipartite entanglement: some examples}
\label{sec:mme}

Let us illustrate this approach on the simplest non-trivial
situation, that of three entangled qubits. If the pure state is
fully factorized, say \beq |\psi\rangle=|k\rangle \eeq for a given
$0\leq k\leq 7$, then the reduced density matrix $\rho_A$ of every
qubit is a pure state, whence \beq
p(N_{AB})=\delta_{N_{AB},1}:\eeq there is no entanglement. On the
other hand, for a maximally entangled state
\begin{equation}\label{eq:psi1}
|\psi\rangle=\frac{1}{\sqrt{2}}(|000_2\rangle+|111_2\rangle) ,
\end{equation}
one gets a completely mixed state for every partition, namely
$\rho_A=\mathrm{I}_2/2$ and thus
\begin{equation}\label{eq:p1}
p(N_{AB})=\delta_{N_{AB},2} ,
\end{equation}
with maximum average and zero variance: there is maximum
multipartite entanglement, fully distributed among the three
qubits. The above probability distributions should be compared
with an intermediate case like
\begin{equation}\label{eq:psi2}
|\psi\rangle=\frac{1}{\sqrt{2}} (|000_2\rangle+|110_2\rangle),
\end{equation}
where the first couple of qubits are maximally entangled (Bell
state) while the third one is completely factorized. In such
situation one gets $\rho_1=\rho_2=\mathrm{I}_2/2$, while
$\rho_3=|1\rangle\langle1|$, whence
\begin{equation}\label{eq:p1bis}
p(N_{AB})=\delta_{N_{AB},1}/3 + 2 \delta_{N_{AB},2}/3 .
\end{equation}
This simple application discloses the rationale behind the quantity
$p(N_{AB})$ as a measure of multipartite entanglement.

When the system becomes larger, the natural extension is towards
larger (balanced) bipartitions. We stress that, besides the
comment that follows Eq.\ (\ref{eq:propNAB}), the use of balanced
bipartitions is simply motivated by the fact that, in the
thermodynamical limit, the unbalanced ones give a small
contribution, from the statistical point of view: this can be
easily understood if one considers that for $n$ large and $n_A \ll
n$ the binomial coefficients
 \beq\label{eq:baldomin}
\left(\begin{array}{c}
  n \\
  n/2 \\
\end{array}\right)\gg \left(\begin{array}{c}
  n \\
  n_A \\
\end{array}\right) ,
\eeq
so that our characterization of multipartite entanglement will be
largely dominated by balanced bipartitions. Notice also that very
unbalanced bipartitions of large systems yield negligible average
entanglement \cite{kendon} \footnote{However, particularly for small
systems, but sometimes also for large systems (see later), whenever
a finer resolution is needed, unbalanced bipartitions can also be
considered.}. For all these reasons, if one considers the
distribution over all bipartitions, the contribution from the
balanced bipartitions will dominate due to (\ref{eq:baldomin}). By
contrast, if only \emph{un}balanced bipartitions are considered the
results will be in general very different.

It is interesting to study the features of the characterization of
entanglement proposed in Sec.\ \ref{sec:syst} when applied to
particular classes of states. For the GHZ states \cite{ghz} we
find
\beq\label{eq:NABGHZ}
 N_{AB}(\text{GHZ})=2
 \eeq
for all possible bipartitions (both balanced and unbalanced) and for
an arbitrary number of qubits. Clearly, the width of the
distribution is 0, i.e.\ $p(N_{AB})=\delta_{N_{AB},2}$.

For the W states \cite{w} we obtain
\beq\label{eq:NABW}
N_{AB}(\text{W})=\frac{n^2}{n_A^2+n_B^2}.
\eeq
This value depends only on the relative size of the two partitions,
i.e.\ also in this case the width of the distribution of bipartite
entanglement is 0. Notice that, if $n$ is even, $N_{AB}(\text{W})=2$
for balanced bipartitions (and in this case a discrimination between
W and GHZ states would require the analysis of unbalanced
bipartitions). Moreover, in the large $n$ limit
$N_{AB}(\text{W})\simeq 2$ also for $n$ odd.

These results indicate that, for $n$ large, the amount of
(multipartite) entanglement is limited both for GHZ and W states.
These states essentially share the same amount of entanglement
when $n$ is large. They can be distinguished only by considering
less relevant (from the statistical point of view) bipartitions.
Moreover, for $n$ large, $N_{AB}(\text{W})\neq 1$ for balanced
bipartitions. This means that also in the thermodynamical limit
the W states retain some entanglement.

\section{Typical states}
\label{sec:typstate}

Let us now study the typical form of our characterization of
multipartite entanglement $p(N_{AB})$ for a very large class of
pure states of the form (\ref{eq:genrandomx}), sampled according
to a given statistical law. Several features of these random
states are already known in the literature \cite{SC,aaa,EH}, but
we shall focus on those quantities that are relevant for our
purpose. We write
\begin{equation}
|\psi\rangle = \sum_{k=0}^{N-1} r_k e^{i\phi_k} |k\rangle ,
\label{eq:genrandomr}
\end{equation}
where $\phi_k$ are independent random variables with expectation
\beq\label{eq:espt}
E[e^{i\phi_k}]=0
\eeq
and ${\bm r}=(r_1,\dots,r_N)$ is a random point with a given
symmetric distribution $p(\bm{r})$ on the hypersphere
$S^{N-1}=\{{\bm r}\in\mathbb{R}^N | {\bm r}^2=1\}$. The features
of these random states are readily evaluated: one first splits
$\pi_{AB}$ in two parts
\begin{equation}\label{eq:randomwalk}
\pi_{AB}=X_{AB}+M_{AB} ,
\end{equation}
where
\begin{eqnarray}\label{eq:XAB}
X_{AB}&=&{\sum_{j,j'}}' {\sum_{l,l'}}' r_{j l} r_{j' l} r_{j' l'}
r_{j l'} e^{i(\phi_{j l}-\phi_{j' l}+\phi_{j' l'}-\phi_{j l'})},\nonumber\\
\\
M_{AB}&=&{\sum_{j,j'}}' {\sum_l} r_{j l}^2 r_{j' l}^2 + \sum_j
{\sum_{l,l'}}' r_{j l}^2 r_{j l'}^2 + \sum_{j,l} r_{jl}^4 ,
\label{eq:MAB}
\end{eqnarray}
with $j,j'=0,\dots,N_A-1$, $l,l'=0,\dots,N_B-1$, and primes banning
equal indices in the sums.

We note that the expectation value $E[r_{jl}^2]=O(1/N)$, thus
$X_{AB}$ and $M_{AB}$ are sums of at most $N^2$ terms of order
$1/N^2$. By the central limit theorem, for large $N$, $\pi_{AB}$
tends to a Gaussian random variable with mean and variance
\barr \mu_{AB}&=& E[\pi_{AB}] , \nonumber\\
\sigma^2_{AB}&=&E[\pi_{AB}^2]-\mu_{AB}^2, \earr respectively,
namely it is distributed as
\begin{equation}\label{eq:Gauss}
f(\pi_{AB})=\frac{1}{(2\pi\sigma_{AB}^2)^{\frac{1}{2}}}
\exp\left(-\frac{(\pi_{AB}-\mu_{AB})^2}{2\sigma_{AB}^2}\right).
\end{equation}
From $E[X_{AB}]=0$ and the independence between phases $\phi_k$ and
moduli $r_k$ we get
\begin{equation}\label{eq:muAB}
\mu_{AB}=E[M_{AB}]=N(N_A+N_B-2)E[r_1^2r_2^2]+N E[r_1^4]
\end{equation}
and
\begin{equation}\label{eq:sigma2AB}
\sigma^2_{AB}=E[X^2_{AB}]+ E[M^2_{AB}] -\mu^2_{AB},
\end{equation}
where
\begin{equation}\label{eq:EX2}
E[X^2_{AB}]=2N(N_A-1)(N_B-1) E[r_1^2 r_2^2 r_3^2 r_4^2]
\end{equation}
and
\begin{eqnarray}
& & E[M^2_{AB}]= N(N_A + N_B-2)\nonumber\\
& & \qquad\quad\times [(N_A+N_B)(N-4)-2(N-5)] E[r_1^2 r_2^2
r_3^2 r_4^2] \nonumber\\
& & + 2N(N_A+N_B-2)(N+2N_A+2N_B-8) E[r_1^2 r_2^2 r_3^4] \nonumber\\
& & + N(N+2N_A+2N_B-5) E[r_1^4 r_2^4] \nonumber\\
& & + 4N(N_A+N_B-2) E[r_1^2 r_2^6] + N E[r_1^8] ,
\label{eq:EM2}
\end{eqnarray}
where we used $E[r_1^\alpha r_2^\beta r_3^\gamma
r_4^\delta]=E[r_i^\alpha r_j^\beta r_l^\gamma r_k^\delta]$ with
$i,j,l,k$ all distinct. Notice that the above results do not
depend on the particular distribution of $\phi_k$, as far as the
condition (\ref{eq:espt}) is satisfied (otherwise the analysis is
still valid, but Eqs.\ (\ref{eq:muAB})-(\ref{eq:EM2}) become more
involved). Our results particularize for the case of a typical
pure state (\ref{eq:genrandomx}), sampled according to the
unitarily invariant Haar measure, where each $z_k\in\mathbb{C}$ is
chosen from an ensemble that is uniformly distributed over the
projective Hilbert space $\sum_k |z_k|^2 =1$. In such a case, in
(\ref{eq:genrandomr}), $\phi_k \in [0,2\pi]$ are independent
uniformly distributed random variables and ${\bm
r}=(r_1,\dots,r_N)$ is a random point uniformly distributed on the
hypersphere $S^{N-1}$, with distribution function
\begin{equation}\label{eq:pr}
p(\bm{r})= \frac{2^{N}}{ \pi^{N/2}} \,
\Gamma\left(\frac{N}{2}\right) \delta(1-\bm{r}^2) ,
\end{equation}
the prefactor being twice the inverse area of the hyperoctant
$\{r_i>0\}$, with $\Gamma(x)$ the Gamma function.

\begin{figure*}
\includegraphics[width=0.7 \textwidth]{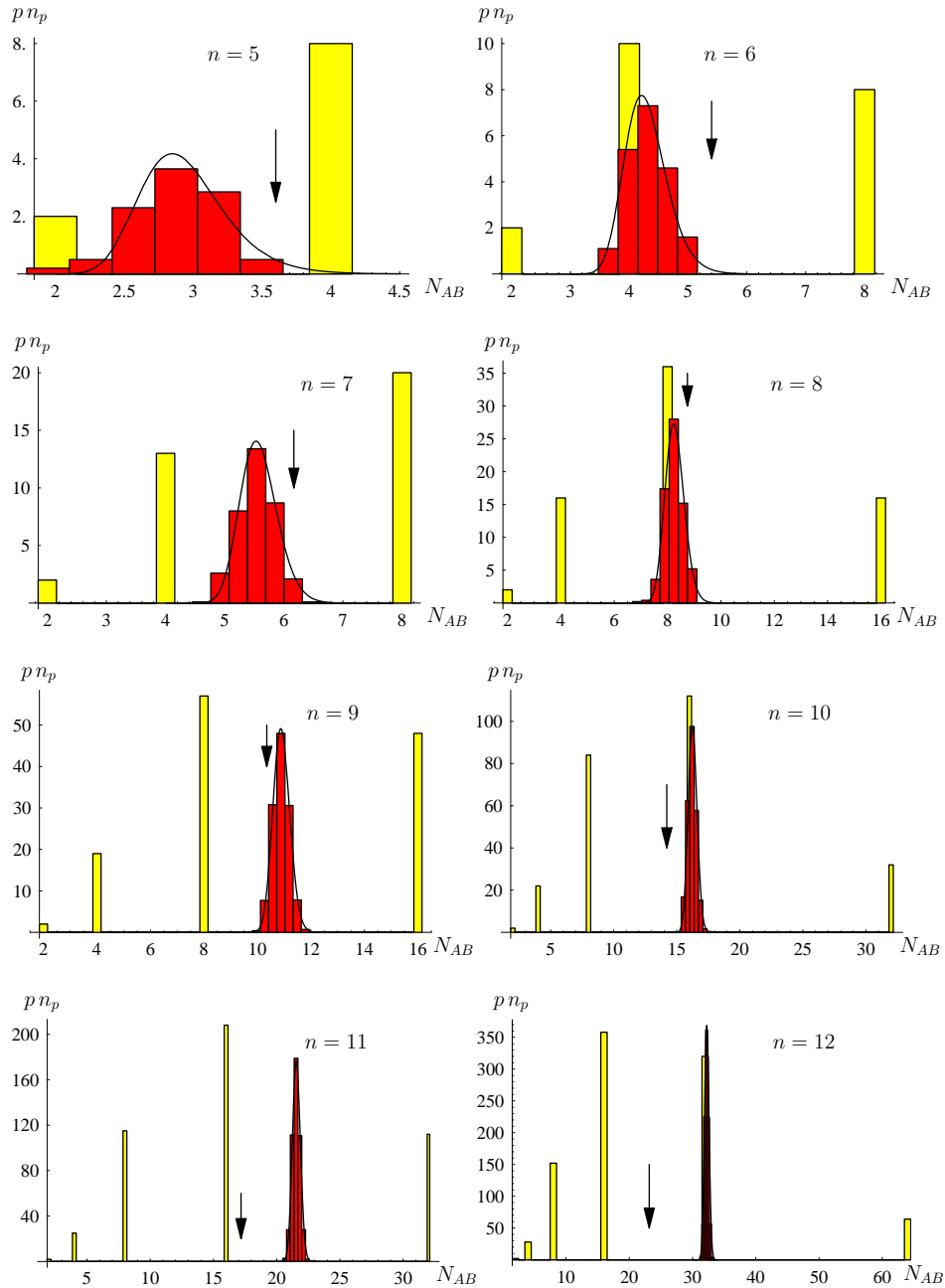}
\caption{Number of balanced bipartitions vs $N_{AB}$; $p$ is the
probability density, $n_p=n!/n_A!n_B!$ is the number of
bipartitions. The yellow bars represent one-dimensional cluster
states [see Eq.\ (\ref{eq:cluster})], the red ones random states;
the solid line is the distribution
(\ref{eq:muABas})-(\ref{eq:distrpartnumber}); the black arrows
indicate the average $\langle N_{AB}\rangle_{\text{cluster}}$. For
even $n$ ($n=12$ in particular) the distribution of the random
state partially hides a peak of the corresponding cluster state
distribution, centered at $N_{AB}= 2^{n_A-1} = 2^{[n/2]-1}$.}
\label{confrontorandomclusternew}
\end{figure*}

The explicit expressions of (\ref{eq:muAB})-(\ref{eq:EM2}) can be
computed through (\ref{eq:pr}), recovering the values of mean and
variance obtained by different approaches \cite{aaa,SC,EH}.
However one can easily estimate them for large $N$ by the
following reasoning. For large $N$ the marginal distributions of
the amplitudes $r_k$ become normal,
\begin{eqnarray}
p(r_k)&=&\frac{2}{\sqrt{\pi}}
\frac{\Gamma(N/2)}{\Gamma\left((N-1)/2\right)}{\left(1-r_k^2\right)}^{(N-3)/2}
\nonumber\\
&\sim& 2 \sqrt{\frac{N}{2\pi}} \exp\left(-\frac{N}{2} r_k^2
\right) \qquad (\forall k) ,
\label{eq:1partdistr}
\end{eqnarray}
with variance $1/N$. One can convince oneself of the correctness of
the above expression just by recalling the asymptotic behavior of
gamma function and expanding $(1-r_k^2)^{N/2}$. Moreover it is not
difficult to show that the $r_k$'s become uncorrelated, hence
independent. Therefore the expectation of products factorizes and
$E[r_1^{2 m}]=(2m-1)!!/N^m$, yielding
\begin{equation}\label{eq:muABas}
\mu_{AB}=\frac{N_A+N_B-1}{N} , \quad \sigma^2_{AB}=\frac{2}{N^2} .
\end{equation}
It is important to notice that when $N\gg 1$ we can effectively
replace $r_k$ with its mean square root value, $r_k=1/\sqrt{N}$,
from which (\ref{eq:muABas}) immediately follows. In the
simulation plotted in Fig.\ \ref{confrontorandomclusternew} we
used the above substitution. The fact that for Haar distributed
states the average (\ref{eq:muABas}) is concentrated around a
large value was already recognized by other authors
\cite{aaa,SC,EH}.
\begin{table}
\begin{tabular}{||c||c|c|c|c||}
\hline
$n$  & GHZ & W & cluster & random  \\
\hline
5 & 2 & 1.923 & 3.6 & 2.909 \\
\hline
6 & 2 & 2 & 5.4 & 4.267 \\
\hline
7 & 2 & 1.96 & 6.171 & 5.565 \\
\hline
8 & 2 & 2 & 8.743 & 8.258 \\
\hline
9 & 2 & 1.976 & 10.349 & 10.894 \\
\hline
10 & 2 & 2 & 14.206 & 16.254 \\
\hline
11 & 2 & 1.984 & 17.176 & 21.558 \\
\hline
12 & 2 & 2 & 23.156 & 32.252\\
\hline
\end{tabular}
\caption{\label{tab:confront} Mean bipartite entanglement
$E[N_{AB}]$, analitically evaluated according to Eqs.\
(\ref{eq:NABGHZ}), (\ref{eq:NABW}) and (\ref{eq:muABas}). The values
for the cluster state were computed by inserting (\ref{eq:cluster})
in the definitions (\ref{eq:NAB})-(\ref{eq:piAB}). }
\end{table}

The quantity of interest is $N_{AB}$ defined in
Eq.~({\ref{eq:NAB}}). From Eq.~(\ref{eq:Gauss}), its probability
density reads
\begin{eqnarray}
p(N_{AB})&=&\frac{1}{N^2_{AB}(2\pi\sigma_{AB}^2)^{1/2}}
\exp\left(-\frac{(N_{AB}^{-1}-\mu_{AB})^2}{2\sigma_{AB}^2}\right).\nonumber\\
\label{eq:distrpartnumber}
\end{eqnarray}

\begin{figure}
\includegraphics[width=0.45 \textwidth]{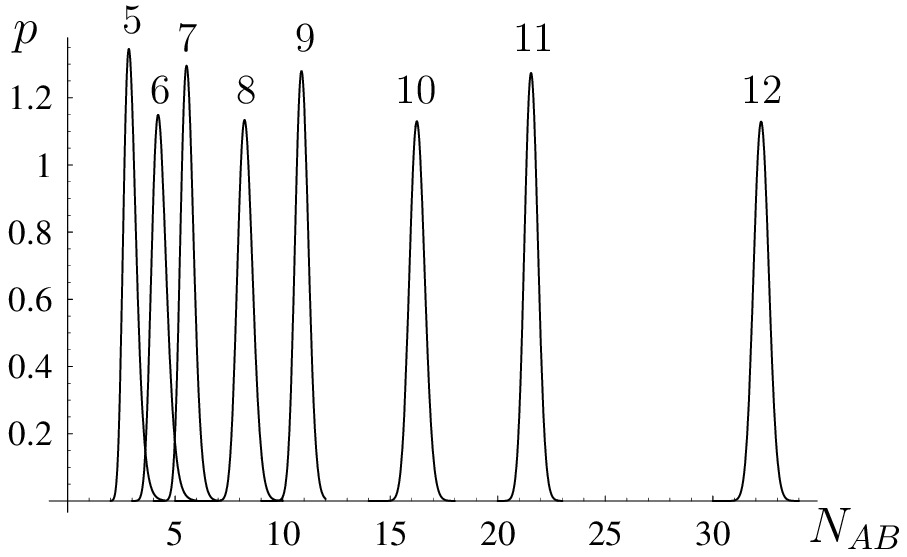}
\caption{Probability densities functions
(\ref{eq:distrpartnumber}) vs $N_{AB}$. Each curve is labeled with
the corresponding value of $n$ (number of qubits). The standard
deviation of the distribution is essentially independent of $n$.}
\label{picprob}
\end{figure}

It is interesting to compare the features of the random states
with those of other states studied in the literature. Table
\ref{tab:confront} displays the average value of $N_{AB}$
(evaluated for $n=5\div12$) for GHZ states \cite{ghz}, W states
\cite{w}, the generic states (\ref{eq:genrandomr}) and
one-dimensional cluster states \cite{briegel} defined as
\begin{equation}
|\phi_n\rangle = \frac{1}{\sqrt{2^n}}\bigotimes_{k=1}^{n} (\ket
0_k \sigma^{(k+1)}_z+\ket 1_k) , \label{eq:cluster}
\end{equation}
where $\sigma_z$ is the third Pauli matrix and the convention
$\sigma^{(n+1)}_z=1$ is applied. While the entanglement of the GHZ
and W states is essentially independent of $n$ [see Eqs.\
(\ref{eq:NABGHZ})-(\ref{eq:NABW})], the situation is drastically
different for cluster and random states. In both cases, the average
entanglement increases with $n$; for $n>8$ the average entanglement
is higher for random states. However, it is now clear that the
average $E[N_{AB}]$ yields poor information on multipartite
entanglement. For this reason, it is useful to analyze the
distribution of bipartite entanglement over all possible balanced
bipartitions. The results for the cluster and random states are
shown in Fig.\ \ref{confrontorandomclusternew}, for $n=5\div 12$,
where the product of the probability density $p$ times the number of
bipartitions $n_p=n!/n_A!n_B!$ is plotted vs $N_{AB}$. Notice that
the distribution function of the random state is always peaked
around $\mu_{AB}^{-1}$ in (\ref{eq:muABas}) and becomes narrower for
larger $n$, in agreement with $\sigma^2_{AB}$ in (\ref{eq:muABas}).
Notice also that the cluster state can reach higher values of
$N_{AB}$ (the maximum possible value being $2^{[n/2]}$), however,
the fraction of bipartitions giving this result becomes smaller for
higher $n$. This is immediately understood if one realizes that
cluster states are designed for optimized applications and therefore
perform better in terms of specific bipartitions. On the other hand,
according to the characterization we propose, the random states
(\ref{eq:genrandomr}) are characterized by a large value of
multipartite entanglement, that is roughly independent on the
bipartition. The probability density functions
(\ref{eq:distrpartnumber}) are displayed in Fig.\ \ref{picprob}.

A few additional comments on  random states are in order. In the
thermodynamical limit
\begin{equation}\label{eq:TDmeasure}
\frac{\sigma_{AB}}{\mu_{AB}} =
\frac{\sqrt{2}}{N_A+N_B-1} = O(1/\sqrt{N})
\end{equation}
and the single real number $E[N_{AB}]$ is sufficient to characterize
multipartite entanglement (modulo more accurate thermodynamical
considerations).

In general, for finite systems, the mean bipartite entanglement
$N_{AB}\simeq\mu_{AB}^{-1}$ in (\ref{eq:muABas}) is maximum for $N_A
= N_B = \sqrt{N}$ ($N_A = N_B/2 = \sqrt{N/2}$) for even (odd) $n$,
namely for balanced bipartitions. Notice however that, as we already
emphasized a number of times in this article, although we focused on
balanced bipartitions for illustrative purposes, the main results
are valid when one includes also unbalanced bipartitions, as, by
virtue of (\ref{eq:baldomin}), the contribution of the balanced
bipartition will be exponentially dominant.

Moreover, for large $N$, any (symmetric) radial distribution $p(\bm
r)$ yields the same results (\ref{eq:muABas}), the only relevant
feature being the curvature in the projective Hilbert space, forced
by the normalization $\bm r^2=1$ [see for example (\ref{eq:pr})]. In
this sense, the above analysis is of general validity, being
independent of the particular choice of the ensemble.

\section{Comparison with some multipartite entanglement measures}
\label{sec:compar}

It is interesting to compare our proposed characterization of
multipartite entanglement with some other entanglement measures. In
general, we will find that this characterization sheds additional
light on this issue and helps specify some of the global features of
multipartite entanglement in a clear-cut way.

The quantity
\cite{brennen}
\beq\label{eq:Q}
 Q(\ket \psi)=2\left(1-\frac{1}{n}\sum_k\mathrm{Tr}{\rho^2_{\{k\}}}\right),
\eeq
where $\rho_{\{k\}}$ is the reduced density matrix of qubit $k$,
i.e.\ $\rho_A$ with $A=\{k\}$. In our language, it corresponds to
the mean value of $\pi_{AB}$ over maximally unbalanced bipartitions,
namely
\beq\label{eq:Q1}
 Q(\ket
 \psi)=2\left(1-E_{\text{max unbal}}[\pi_{AB}]\right).
\eeq
For W states this yields $Q(\text{W})\sim 0$ for large $n$. This
should be compared with the value $N_{AB}(W)=2$ (exact for even $n$,
approximate for odd $n$), obtained by considering balanced
bipartitions of the system. As previously stressed, this means that
the W states retain some entanglement even in the thermodynamical
limit.

Moreover, at variance with $Q$, the mean value of $N_{AB}$ can
distinguish sub-global entanglement. For instance, the state
$\ket\psi=(\ket0\ket0+\ket1\ket1)\otimes (\ket0\ket0+\ket1\ket1)/2$
cannot be distinguished from the GHZ state by using only $Q$. On the
other hand, one gets an average $\langle N_{AB}\rangle=3$ and a
width for the distribution $\sigma=1.55$. Another interesting point
is that the distribution of $N_{AB}$ can distinguish GHZ and cluster
states (actually the average is already sufficient, as can be seen
from Table
\ref{tab:confront}). From these results one can argue that the
probability density function of the participation number $N_{AB}$
not only better specifies the meaning of $Q$ but also yields
additional information.

It is also interesting to recall the behavior of the pairwise
entanglement (concurrence) and the tangle \cite{multipart}. The
former is defined (for states $\rho_{\{i,j\}}$ of two qubits $i$ and
$j$) as
\beq\label{eq:conc}
C_{ij}=\mbox{max}(0,\lambda_1-\lambda_2-\lambda_3-\lambda_4) ,
\eeq
where $\lambda_k$ are the square roots of the eigenvalues (in
decreasing order) of the matrix
$\rho_{\{i,j\}}\sigma_y\otimes\sigma_y\rho^*_{\{i,j\}}\sigma_y\otimes\sigma_y$,
and is therefore related to $\pi_{AB}$ with $A=\{i,j\}$ (highly
unbalanced bipartitions when $N$ is large). The tangle is defined as
\begin{equation}\label{eq:tangle}
\tau_1^{(i)}=4 \det \rho_{\{i\}}=2(1-\mathrm{Tr}{\rho^2_{\{i\}}}),
\end{equation}
where $\rho_{\{i\}}$ is the reduced density matrix for qubit $i$.
Note that $\tau_1^{(i)}= 2(1-\pi_{AB})$, with $A=\{i\}$, is nothing
but the local version of $Q$ in (\ref{eq:Q}). In particular one can
consider the ratio $R^{(i)}=\tau_2^{(i)}/\tau_1^{(i)}$
\cite{tognetti} where $\tau_2^{(i)}=\sum_{j\neq i}C_{ij}^2$ is the
sum of the squared concurrences of qubit $i$ with qubit $j$. Due to
the Coffman-Kundu-Wootters conjecture $\tau_1^{(i)}\geq
\tau_2^{(i)}$ \cite{multipart} one can take $R^{(i)}$ as a witness
of multipartite entanglement: if $R^{(i)}<1$ pairwise entanglement
is less relevant than multi-qubit correlations. In particular, in
order to elucidate their relation with the bipartite entanglement of
highly unbalanced bipartitions, it is interesting to apply these
measures to typical states. We notice that, in the limit of large
$n$ one has, on the average,
\begin{eqnarray}\label{eq:tangletypical}
E[\tau_1]&=& Q = 1-1/2^{n-1}\sim1, \nonumber \\
E[\tau_2]&\sim& 0.
\end{eqnarray}
These results are interesting because they show how, in the
thermodynamical limit, pairwise entanglement is negligible for
typical states. At the same time, Eq.\ (\ref{eq:tangletypical}) does
not yield much information about the very structure of multipartite
entanglement: actually one can see that the same result can be
obtained for GHZ states (for arbitrary $n$). In this sense our
characterization in terms of the probability density function
corroborates and better specifies the results obtained by studying
the behavior of $R$.



\section{Conclusions}
\label{sec:conc}

It is well known  that an efficient way to generate states endowed
with random features is by a chaotic dynamics
\cite{entvschaos,SC}, or at the onset of a quantum phase
transition \cite{QPT}. In particular, the random states
(\ref{eq:genrandomr}) describe quite well states with support on
chaotic regions of phase space, before dynamical localization has
taken place. Interestingly, other ways have been recently proposed
\cite{EH,plenio} in order to generate these states, in particular
by operating on couples of qubits with random unitaries followed by
CNOT gates \cite{plenio}. The introduction of a probability density
function as a measure of multipartite entanglement paves the way to
further investigations of this intimate relation between
entanglement and randomness. Work is in progress in order to clarify
whether the random states can be efficiently used in quantum
information processing.

In some sense, the characterization we propose quantifies the
robustness of entanglement against all possible partial tracing.
Clearly, it is more effective for large number of qubits and when
relatively few moments are sufficient to specify the distribution.
We stress that although we studied the distribution function of the
inverse purity (linear entropy) (\ref{eq:NAB}), our analysis could
have been performed in terms of any other measure of bipartite
entanglement, such as the entropy.

Finally, we emphasize again the main motivation behind this work:
as the number of subsystems increases, the number of measures
(i.e.\ real numbers) needed to quantify multipartite entanglement
grows exponentially. It is therefore not surprising if a
satisfactory global characterization of entanglement requires the
use of a function.

\acknowledgments
This work is partly supported by the bilateral Italian--Japanese
Projects II04C1AF4E on ``Quantum Information, Computation and
Communication'' of the Italian Ministry of Instruction, University
and Research and by the European Community through the Integrated
Project EuroSQIP. G.F.\ acknowledges the support and kind
hospitality of the Department of Physics of Waseda University, where
part of this work was done.


\end{document}